\documentclass[12pt,a4paper]{article}

\usepackage[british]{babel}

\usepackage[a4paper,top=2cm,bottom=2cm,left=2.5cm,right=2.5cm,marginparwidth=1.75cm]{geometry}
\usepackage{amssymb}

\usepackage[colorlinks=true,urlcolor=blue]{hyperref}

\newcommand{\hepbook}[3]{%
  #1, \textit{#2} (#3).%
}

\newcommand{\heparticle}[6]{%
  #1, #2, #3 #4 (#5) #6%
}

\usepackage{amsmath}
\usepackage{graphicx}
\usepackage{orcidlink}
\usepackage[title]{appendix}
\usepackage{mathrsfs}
\usepackage{amsfonts}
\usepackage{booktabs} % For \toprule, \midrule, \botrule
\usepackage{caption}  % For \caption
\usepackage{threeparttable} % For table footnotes
\usepackage{algorithm}
\usepackage{algorithmicx}
\usepackage{algpseudocode}
\usepackage{listings}
\usepackage{enumitem}
\usepackage{chngcntr}
\usepackage{booktabs}
\usepackage{lipsum}
\usepackage{subcaption}
\usepackage{authblk}
\usepackage[T1]{fontenc}    % Font encoding
\usepackage{csquotes}       % Include csquotes
\usepackage{diagbox}
\usepackage{comment}

\usepackage{setspace}
\usepackage{titlesec}
\titleformat{\section} % Redefine section numbering format
  {\normalfont\Large\bfseries}{\thesection.}{1em}{}
  
\usepackage{lineno} % Add the lineno package

\rightlinenumbers % Right-align line numbers

\linenumbers % Enable line numbering

\newcommand{\subsubsubsection}[1]{%
  \vspace{\baselineskip}% Add some space
  \noindent\textbf{#1\\}\quad% Adjust formatting as needed
}
\usepackage{float}   % for customizing caption position
\usepackage{caption} % for customizing caption format


\nolinenumbers
\title{Constrained Dynamics on an Ellipse}

\author[1]{Akshay Chaturvedi}
\author[2]{Pichai Ramadevi}
\affil[1, 2]{\small Department of Physics, Indian institute of Technology Bombay, Mumbai 400076, India}

\begin{document}
\maketitle

\begin{abstract}
We first review the application of Dirac’s method to the dynamics of a classical particle constrained to a circle and its subsequent quantization. Then, we extend the analysis to a particle constrained to move on an ellipse. Particularly, we identify the corresponding Dirac brackets and determine the quantum operators associated with the fundamental dynamical variables. 
%We also comment on alternative approaches to quantizing such constrained systems, although these methods do not immediately yield the final operator expressions.
\end{abstract}


%-------------------------------------------
% Paper Body
%-------------------------------------------
%--- Section ---%
\begin{comment}
\section*{Nomenclature}

\begin{tabbing}
$T$\qquad \= Temperature (K)\\
$u_i$ \> Velocity in the x-direction (m/s)\\
$\tau_{ij}$ \> Shear stress (N/m2)\\
$\omega$ \> Specific turbulent dissipation rate (1/s)\\
$Y_\omega$ \> Dissipation of $\omega$
\end{tabbing}
\end{comment}

\pagebreak
%--- Section ---%
\section{Introduction}

A standard approach to quantize classical dynamical systems is to express the theory in Hamiltonian form and promote the phase–space variables to operators satisfying canonical commutation relations. 
For constrained systems, the Dirac’s method \cite{Dirac1958} provides a systematic framework for extending the procedure. Note that the Lagrangian for such systems is singular. That is, the conjugate momenta cannot be inverted to express the velocities in terms of the coordinates and momenta

The analysis in Ref.\cite{Scardicchio2002} examines the quantization of a particle constrained to move on a circle using two approaches. 

$\bullet$ The first is a straightforward procedure working with the reduced variables appropriate to the constraint. For circle, the angular coordinate $\theta$ as the reduced coordinate is sufficient to deduce the conjugate momenta $p_{\theta}$. 
%If such a choice of reduced coordinates are known for a constrained system , then  Dirac’s formalism setting is not required.
For a general constrained system,  such reduced variables may not be manifest requiring the 
the second approach.

$\bullet$ The second approach employs Dirac’s method, in which the full set of constraints is identified and the Dirac brackets are computed to determine the corresponding operator algebra. 
Interestingly, this approach applied to the circle showed some structural features  of the system that were invisible in the previous approach.

As emphasized in \cite{Scardicchio2002}, second approach can capture information on the topology of the problem and reveal an an additional term in the Hamiltonian. Such a term from Dirac bracket computation can be interpreted as the contribution from the constraining mechanism. These insights constitute an advance over the first approach.

%“in this process … the information on the topology of the problem is made explicit,” and the Dirac-bracket computation reveals an additional term in the Hamiltonian interpreted as the contribution from the constraining mechanism. 

Our focus is to extend this analysis to a particle constrained to move on an ellipse. Clearly, there are additional difficulties due to the absence of rotational symmetry on the ellipse. In particular, several features that rely on the special geometry of “circular” (trigonometric) functions have no direct counterpart for an ellipse. Hence, heuristically extrapolating these relations fails to produce the correct operator algebra. As a result, even within Dirac’s framework, obtaining the operator expressions requires a more involved and explicit computation.

%In extending this analysis to a particle constrained to an ellipse, additional difficulties arise that have no analogue in the circular case. The loss of rotational symmetry eliminates the identities and simplifications that make many of the intermediate steps essentially elementary for the circle. In particular, several features that rely on the special geometry of “circular” (trigonometric) functions have no direct counterpart for an ellipse, and heuristically extrapolating these relations fails to produce the correct operator algebra. As a result, even within Dirac’s framework, obtaining the operator expressions requires a more involved and explicit computation. In what follows, 
The plan of the paper is as follows: In section \ref{sec2}, we will review the salient aspects of Dirac's method for constrained systems. We present the application of the Dirac's method for the motion of a particle on an ellipse in section \ref{sec3}.
Specifically, we highlight some of the key differences by comparing the circle with the ellipse.
In the concluding section \ref{sec4}, we will summarize our results.

\section{The Dirac Method}\label{sec2}
We will briefly present the essential tools involved in the Dirac's method to tackle constrained system\cite{Sudarshan1979}.
Consider a Lagrangian $L(\{x_i\},\{\dot{x_i}\})$ for a system described by $N$ coordinates $x_1,x_2, \ldots x_N$. The classical
dynamics follows from the principle of stationary action:
\begin{equation}
    S[\{x_i\}] = \int_{t_0}^{t_1} dt\, L(\{x_i\},\{\dot{x_i}\}), 
    \qquad 
    \delta S[\{x_i\}] = 0,
\end{equation}
which yields the $N$ Euler-Lagrange equations
\[
\frac{\partial L}{\partial x_i}
- \frac{d}{dt}\left( \frac{\partial L}{\partial \dot{x}_i} \right)
= 0~~\forall~ \{x_i\}
\]
The canonical momenta and Hamiltonian are defined in the usual way:
\begin{equation}
    p_i = \frac{\partial L}{\partial \dot{x}_i};~~
    H(\{x_i\},\{p_i\}) = \sum_i p_i\, \dot{x}_i- L(\{x_i\},\{\dot{x_i}\});~~ 
    \{x_i,p_j\}_{PB} = \delta_{ij}.
\label{eq:momenta1}
\end{equation}
The time-evolution of any phase-space function or dynamical variable $A(\{x_i\},\{p_i\})$ with no explicit time dependence is given by
\[
\dot A = \{A,H\}_{PB}~.
\]
Recall that the Legendre transform leading to writing Hamiltonian from the Lagrangian is possible only if the relations $p_j(x,\dot{x})$ can be inverted to give $\dot{x}_j(x,p)$. If the inversion fails (even locally), the
Lagrangian is \emph{singular}: its Hessian with respect to the velocities has
vanishing determinant,
\[
\det\!\left( \frac{\partial^2 L}{\partial \dot x_i\,\partial \dot x_j} \right) = 0.
\]
%Two situations can occur. If the relations $p_j(x,\dot{x})$ can be inverted %to give
%$\dot{x}_j(x,p)$, the Legendre transform is well defined and the standard
%Hamiltonian formulation applies. If the inversion fails (even locally), the
%Lagrangian is \emph{singular}: its Hessian with respect to the velocities has
%%vanishing determinant,
%\[
%\det\!\left( \frac{\partial^2 L}{\partial \dot x_i\,\partial \dot x_j} %\right) = 0.
%\]
For such singular Lagrangians, where the defining relations of some of the  momenta\eqref{eq:momenta1} are not independent, we can obtain $\phi(\{x_i\},\{p_i\}) \approx 0$  
which  must be regarded as
\emph{constraints}. 
There could be more than one independent relation. Suppose there are $M$ constraints:
\begin{equation}
\phi_i \approx 0, ~{\rm where}~i\in [1,2,\ldots M]~.\label{constra}
\end{equation}
Note that the constraints vanish only \emph{weakly}. That is, they may have non-vanishing Poisson
brackets with other phase-space functions. These $M$ constraints, arising
directly from the definition of the momenta, are termed \emph{primary}
constraints.

Our naive Hamiltonian formalism breaks down for constrained systems.
We need to implement Dirac's prescription incorporating the above constraints as follows:
%A Hamiltonian must ultimately be a function of $x$ and $p$ only. When the
%velocity dependence cannot be eliminated, the naïve Hamiltonian formalism
%breaks down. Dirac’s prescription is to introduce the \emph{total Hamiltonian},
\[
H_T = H + \sum_{j=1}^M u_j(t)\,\phi_j(\{x_i\},\{p_i\}),
\]
where $H_T$ is the total Hamiltonian and  $u_j(t)$'s  are arbitrary functions. For consistency of the eqn.\eqref{constra} satisfied initially, we require the time-evolution to obey:
\begin{equation}
\dot{\phi}_j = \{\phi_j, H_T\}_{PB} \approx 0,
\qquad j = 1,\dots,M.
\label{eq:consistency}
\end{equation}
Enforcing these conditions leads to three possibilities:\\  
(i) trivial equation  \\
(ii) an equation determining some of the multipliers $u_j$\\  
(iii) new relations\\
The new relations give additional constraints called \emph{secondary} constraints.
Then we iterate the evolution of such secondary constraints to tertiary constraints and so on.
Eventually, the procedure terminates when no further constraints arise. This collection of primary, secondary,tertiary constraints and so on, will constitute a set of $J$ constraints.

Recall, these constraints are functions $\phi_j (\{x_i\}, \{p_i\})$ satisfying
\begin{equation}
\phi_j \approx 0,
\qquad
\{\phi_j, H_T\}_{PB} \approx 0.
\label{eq:consistency3}
\end{equation}
The set of all constraints forms a linear space. We can further classify the set into :\\
$\bullet$ \emph{first-class} : which have vanishing Poisson brackets
with \emph{all} other constraints\\
$\bullet$ \emph{second-class} constraints : the remaining constraints. \\
We may find some linear combinations of second-class constraints to yield first-class ones.

For the problem we are focusing, we observe that there are only second-class constraints. Hence we do not discuss the first-class case (see Chapters 8 and 9 of \cite{Sudarshan1979} for a
comprehensive treatment). .

For the constrained system with only  second-class constraints  present, 
we introduce a new canonical brackets - the \emph{Dirac brackets}. 
Define the matrix
\[
M_{ij} = \{\phi_i,\phi_j\}_{PB},
\qquad
G_{ij} = (M^{-1})_{ij},
\]
where the invertibility of $M$ reflects the absence of first-class constraints.
Then the Dirac bracket between any two phase-space functions $A$ and $B$ is
\[
\{A,B\}_D 
= \{A,B\}_{PB}
 - \sum_{i,j=1}^K \{A,\phi_i\}_{PB}\, G_{ij}\, \{\phi_j, B\}_{PB}.
\]
These brackets satisfy all the properties of Poisson brackets, with the added
feature that
\begin{equation}
\{A,\phi_i\}_D = 0,
\qquad
\dot{A} = \{A,H_T\}_{PB} \approx \{A,H_T\}_D,
\end{equation}
the latter following from the consistency conditions.

The Dirac bracket effectively projects the dynamics onto the constraint surface:
it isolates the true physical degrees of freedom of the system. Once the Dirac
brackets have been determined, quantization proceeds by promoting the dynamical
variables to self-adjoint operators and imposing the correspondence principle,
\[
[A,B] \;\longrightarrow\; i\hbar\, \{A,B\}_D.
\]
We now apply this framework to the particle constrained to move on an ellipse and circle in the following section.
%--- Section ---%
\section{Circle versus Ellipse Constrained system}\label{sec3}
We will present the computational steps for particle moving on a circle alongside motion on an ellipse. This will help us clearly notice the differences and complications that arise when trying to solve for the elliptical constraint dynamics by simply extrapolating results from the more manageable case of the circle. 

We will now do a warm up for the simplest circle before we delve on to the ellipse.
\subsection{The Standard Approach}
For motion on a simple  circle or on a sphere, it is easier to change the cartesian coordinates to spherical polar coordinates restricting the radial coordinate to be constant. Thus, we can 
choose the generalized coordinate as $\theta$ for the particle on a circle. This  gives us the standard Euler-Lagrange equations in the $\theta$ coordinate :
\[\frac{d}{dt}\frac{\partial L}{\partial \dot{\theta}} = \frac{\partial L}{\partial \theta}~,\] where $L = \frac{1}{2}r^2\dot{\theta}^2$, where $r$ is a constant.
From the Lagrangian, the Hamiltonian is obtained as follows -
\[H = p_{\theta}\dot{\theta} - L~,\]
and the fundamental bracket (here, the Poisson bracket) comes out to be -
\[ \{\theta, p_{\theta}\}_{PB} = 1~.\]
For quantizing the system, we just map the Poisson brackets to commutators by defining two self adjoint operators $\hat{\theta}$ and $\hat{p}_{\theta}$ and requiring - 
\[ [\hat{\theta}, \hat{p}_{\theta}] = i\hbar~.\]

We can find such a couple of self-adjoint operators in
the Hilbert space $\mathcal{H} = L^2(0, 2\pi)$ and their expression is:

\begin{equation}
    \begin{aligned}       
        \hat{\theta}\psi(\theta)=&\theta\psi(\theta),\\
        \hat{p}_\theta\psi(\theta)=&\hbar\left(-i\frac{\partial}{\partial
        \theta}-\alpha\right)\psi(\theta). 
    \end{aligned}
\end{equation}

We add the constant $\alpha$ in the momentum $p_\theta$ to mimic
the possible presence of a magnetic field enclosed in the circle. Their domains are chosen to be respectively $D_\theta=\mathcal{H}$ and
$D_{p_\theta}=\{\psi\in\mathcal{H}|\psi(0)=\psi(2\pi),\psi'\in \mathcal{H}\}$.
%Notice that we have chosen one of the infinite self-adjoint extensions of the momentum $\hat{p}_\theta$. 
The Hamiltonian reads -
\[
\label{eq:standhamilt}
H(\hat{\theta},\hat{p}_{\theta})=\frac{\hat{p}_{\theta}
^2}{2r_0^2}=\frac{\hbar^2}{2r_0^2}\left(-i\frac{\partial}{\partial
\theta}-\alpha\right)^2, \] and is self-adjoint in the domain of
$p_\theta$, i.e.\ $D_{p_\theta}$. The Schrodinger equation is
(reinserting $m$ and $\hbar$) - 
\[ 
i\hbar\frac{\partial \psi}{\partial
t}=\frac{\hbar^2}{2mr_0^2}\left(
-i\frac{\partial}{\partial\theta}-\alpha \right)^2\psi. \] 
% This is what we expected.
We can draw parallel between the cyclic coordinate $\theta$ (which measures circumference on the circle) with an appropriate cyclic coordinate $K$ (which measures circumference on the ellipse). We will see that $K$ is linked to the elliptic function, as opposed to circular/trigonometric functions.

To obtain $K$, we first impose the ellipse-parameterization on the $x$ and $y$ coordinates (i.e, expressions for $x$ and $y$ such that they automatically satisfy $\frac{x^2}{a^2} + \frac{y^2}{b^2} = 1$) as - 
\[ x = a cos(w)~~;~~y = b sin(w)~, \]
where $a$ and $b$ are constants (lengths of the semi-major and semi-minor axis) of the ellipse respectively). This gives the Lagrangian as -
\[L = \frac{1}{2}\dot{w}^2(a^2sin^2(w) + b^2cos^2(w))\]
Now we propose the following form for $L$ in terms of $K$ :
\begin{equation}
L=\frac{1}{2} (\dot K)^2 = \frac{1}{2}\dot{w}^2[a^2sin^2(w) + b^2cos^2(w)]~,
\end{equation}
implying that $K$ is a cyclic coordinate. Relating this to the above expression, we get \footnote{Choosing the positive square root is valid, as we can interpret $K$ as the distance covered on the circumference, traveling clockwise.},
\begin{equation}
    \begin{aligned}
        &\dot{K} = \dot{w}\sqrt{a^2sin^2(w) + b^2cos^2(w)} \\         
        &K = \int_{0}^{w} dW \sqrt{a^2sin^2(W) + b^2cos^2(W)} \\
        &= b\int_{0}^{w} dW \sqrt{1 - \left(1 - \frac{a^2}{b^2}\right)} \\
        &= b E\left(w, \sqrt{1 - \frac{a^2}{b^2}}\right)
    \end{aligned}
\end{equation}
where $E$ is the elliptic function of the 2nd kind. Given $x$ and $y$, we know $w$ and hence $K$ can be determined. Conversely, $w$ can be obtained in terms of $K$ by inverting the elliptic function. 

%As an aside, it should be mentioned that there was some motivation to use the parameterization $x = Ra cos(w)$, $y = Rb sin(w)$, where $R$ would be a constant, so as to draw a direct parallel with the case of the circle, including the scaling factor (i.e $r$, for the circle) this time, as it seemed it would help later in extrapolating the solutions from quantization of the circle's Dirac brackets. But it turned out to not be of any help, so the usage of $R$ was dropped.

%--- Section ---%
\subsection{Dirac's Approach}
We will explicitly do the analysis for the the case of the ellipse using the Dirac's approach.
In order to compare our results with that of the  circle, we will write down the relevant results in
Ref.\cite{Scardicchio2002}. 

\subsubsection{Constraint Structure and Hamiltonian Analysis}

We now analyse our system using Dirac's procedure for constrained
Lagrangians. For a particle confined to an ellipse,
the Lagrangian is
\begin{equation}
    L(x,\dot{x},y,\dot{y},\lambda)
    = \frac{1}{2}\dot{x}^2 + \frac{1}{2}\dot{y}^2 
    - \lambda\!\left( \frac{x^2}{a^2} + \frac{y^2}{b^2} - 1 \right),
\end{equation}
where $\lambda$ enforces the holonomic constraint.

The canonical momenta follow immediately:
\begin{equation}
    p_x = \dot{x}, \qquad 
    p_y = \dot{y}, \qquad 
    p_\lambda = 0.
\end{equation}
Since $p_\lambda$ cannot be inverted to obtain $\dot{\lambda}$, it is a
\textit{primary constraint},
\begin{equation}
    \phi_1 = p_\lambda \approx 0.
\end{equation}

The canonical Hamiltonian is
\begin{equation}
    H = \frac{p_x^2}{2} + \frac{p_y^2}{2}
        + \lambda\!\left( \frac{x^2}{a^2} + \frac{y^2}{b^2} - 1 \right),
\end{equation}
and the total Hamiltonian is obtained by adding the primary constraint
with an undetermined multiplier $u_1(t)$,
\begin{equation}
    H_T = H + u_1\,\phi_1.
\end{equation}

We now impose the consistency conditions
$\dot{\phi}_j=[\phi_j,H_T]\approx 0$ to determine
secondary constraints. First,
\begin{equation}
    \dot{\phi}_1 
    = -\!\left(\frac{x^2}{a^2} + \frac{y^2}{b^2} - 1\right)
    \approx 0,
\end{equation}
which gives the holonomic constraint as
\begin{equation}
    \phi_2 = \frac{x^2}{a^2} + \frac{y^2}{b^2} - 1 \approx 0.
\end{equation}

Requiring $\dot{\phi}_2\approx 0$ yields
\begin{equation}
    \phi_3 = \frac{x\,p_x}{a^2} + \frac{y\,p_y}{b^2} \approx 0.
\end{equation}

Finally, consistency of $\phi_3$ produces
\begin{equation}
    \phi_4 
    = \frac{p_x^2}{a^2} + \frac{p_y^2}{b^2}
      - 2\lambda\!\left( \frac{x^2}{a^4} + \frac{y^2}{b^4} \right)
    \approx 0.
\end{equation}
The next consistency condition fixes $u_1$ and introduces no further
constraints. Since $p_\lambda$ and $\lambda$ are eliminated by the
Dirac bracket construction, the total Hamiltonian effectively reduces to
\begin{equation}
    H_T = \frac{1}{2}(p_x^2 + p_y^2),
\end{equation}
which is the free Hamiltonian restricted to the constrained surface.

\subsubsection*{Comparison with the Circular Case}

For reference, the circle constraint 
$x^2 + y^2 = r_0^2$ leads to an analogous structure:
\begin{equation}
    \phi_1 = p_\lambda \approx 0,\qquad
    \phi_2=x^2+y^2-r_0^2\approx 0,
\end{equation}
\begin{equation}
    \phi_3 = x p_x + y p_y \approx 0,\qquad
    \phi_4=p_x^2+p_y^2-2\lambda(x^2+y^2)\approx 0.
\end{equation}
The total Hamiltonian again reduces to the free one,
$H_T = \tfrac{1}{2}(p_x^2+p_y^2)$.

The key difference between the two systems does not lie in the constraint
algebra up to this stage - which is structurally similar -but in the shape
of the constraint manifold. For the ellipse, the subsequent Dirac
brackets involve nontrivial denominators originating from its curvature
profile, and this substantially complicates the operator realization in
the quantum theory.

\subsubsection{Constraint Matrix and Dirac Brackets}

With the full set of second-class constraints
$\{\phi_1,\phi_2,\phi_3,\phi_4\}$ obtained, we now construct the
constraint matrix
\begin{equation}
    M_{ij} = \{\phi_i,\phi_j\}_{PB}.
\end{equation}
For the ellipse this takes the form
\begin{equation}
M_{ij} =
\begin{pmatrix}
    0 & 0 & 0 &
    2\!\left(\dfrac{x^2}{a^4}+\dfrac{y^2}{b^4}\right) \\[6pt]
    0 & 0 &
    2\!\left(\dfrac{x^2}{a^4}+\dfrac{y^2}{b^4}\right) &
    4\!\left(\dfrac{x p_x}{a^4}+\dfrac{y p_y}{b^4}\right) \\[6pt]
    0 & -2\!\left(\dfrac{x^2}{a^4}+\dfrac{y^2}{b^4}\right) & 0 &
    2\!\left(\dfrac{p_x^2}{a^4}+\dfrac{p_y^2}{b^4}\right) \\[6pt]
    -2\!\left(\dfrac{x^2}{a^4}+\dfrac{y^2}{b^4}\right) &
    -4\!\left(\dfrac{x p_x}{a^4}+\dfrac{y p_y}{b^4}\right) &
    -2\!\left(\dfrac{p_x^2}{a^4}+\dfrac{p_y^2}{b^4}\right) & 0
\end{pmatrix}.
\end{equation}

It is useful to denote
\begin{equation}
    \alpha = 2\!\left(\frac{x^2}{a^4} + \frac{y^2}{b^4}\right), \qquad
    \beta  = 4\!\left(\frac{x p_x}{a^4} + \frac{y p_y}{b^4}\right), \qquad
    \gamma = 2\!\left(\frac{p_x^2}{a^4} + \frac{p_y^2}{b^4}\right),
\end{equation}
in terms of which $M_{ij}$ simplifies to
\begin{equation}
M=
\begin{pmatrix}
    0 & 0 & 0 & \alpha \\
    0 & 0 & \alpha & \beta \\
    0 & -\alpha & 0 & -\beta \\
    -\alpha & -\beta & -\gamma & 0
\end{pmatrix}.
\end{equation}

The inverse matrix $G=M^{-1}$ is
\begin{equation}
G=
\begin{pmatrix}
    0 & -\dfrac{\gamma}{\alpha^{2}} & \dfrac{\beta}{\alpha^{2}} &
    -\dfrac{1}{\alpha} \\[6pt]
    \dfrac{\gamma}{\alpha^{2}} & 0 & -\dfrac{1}{\alpha} & 0 \\[6pt]
    -\dfrac{\beta}{\alpha^{2}} & \dfrac{1}{\alpha} & 0 & 0 \\[6pt]
    \dfrac{1}{\alpha} & 0 & 0 & 0
\end{pmatrix}.
\end{equation}

For comparison, the circular case
$x^2+y^2=r_0^2$ leads to
\begin{equation}
M_{ij} =
\begin{pmatrix}
    0 & 0 & 0 & 2r_0^2 \\
    0 & 0 & 2r_0^2 & 4\,\mathbf{p}\!\cdot\!\mathbf{r} \\
    0 & -2r_0^2 & 0 & 2\mathbf{p}^2 + 4\lambda r_0^2 \\
    -2r_0^2 & -4\,\mathbf{p}\!\cdot\!\mathbf{r} &
    -2\mathbf{p}^2 - 4\lambda r_0^2 & 0
\end{pmatrix},
\end{equation}
whose inverse may be taken from the standard result in the original literature.

\subsubsubsection{Dirac Brackets}

The Dirac bracket between two functions $A$ and $B$ is
\begin{equation}
\{A,B\}_D = \{A,B\} - \sum_{i,j=1}^4 \{A,\phi_i\}_{PB}\,
G_{ij}\,\{\phi_j,B\}_{PB}.
\end{equation}

Using the inverted matrix above and imposing the elliptical
parameterisation
\[
x=a\cos w,\qquad y=b\sin w,
\]
we obtain:
\begin{equation}
\label{DiracBrac}
\begin{aligned}
\{x,p_x\}_D &= 
i\left(1 - \frac{x^2/a^4}{x^2/a^4 + y^2/b^4}\right)
           = i\,\frac{a^2 \sin^2 w}{a^2 \sin^2 w + b^2 \cos^2 w}, \\[6pt]
\{y,p_y\}_D &= 
i\left(1 - \frac{y^2/b^4}{x^2/a^4 + y^2/b^4}\right)
           = i\,\frac{b^2 \cos^2 w}{a^2 \sin^2 w + b^2 \cos^2 w}, \\[6pt]
\{x,p_y\}_D &= \{y,p_x\}_D
           = -i\,\frac{ab\,\cos w\,\sin w}
                     {a^2\sin^2 w + b^2\cos^2 w}, \\[6pt]
\{p_x,p_y\}_D &= 
 -\,i\,\frac{x p_y - y p_x}{a^2 b^2}
 \left/ \left(\frac{x^2}{a^4} + \frac{y^2}{b^4}\right)\right..
\end{aligned}
\end{equation}

For the circle $x=r_0\cos\theta$, $y=r_0\sin\theta$, the corresponding expressions simplify to
\begin{equation}
\begin{aligned}
[x,p_x]_D &= i\!\left(1-\frac{x^2}{r_0^2}\right), \qquad
[y,p_y]_D = i\!\left(1-\frac{y^2}{r_0^2}\right),\\[4pt]
[x,p_y]_D &= [y,p_x]_D = -\,i\,\frac{xy}{r_0^2},\\[4pt]
[p_x,p_y]_D &= -\,\frac{i}{r_0^2}(x p_y - y p_x),\\[4pt]
[x,y]_D &= 0.
\end{aligned}
\end{equation}

The comparison highlights the source of the analytical difficulty:
for the ellipse, each bracket contains the denominator 
$a^2\sin^2 w + b^2\cos^2 w$,
which has no analogue in the circular case. This difference is ultimately
what prevents a straightforward extrapolation of the circular operator
expressions to the elliptical geometry.

\subsubsubsection{Why the Circular Case Cannot Be Directly Extrapolated}

At this stage, the contrast with the circular constraint becomes clear algebraically. In the circular case, all Dirac brackets reduce
to expressions containing only homogeneous polynomials of $x$, $y$,
$p_x$ and $p_y$, with the single geometric invariant $r_0$ appearing
as an overall scale. This homogeneity makes the operator
realisation essentially unique up to ordering choices, and it is
precisely what allows $L_z = x p_y - y p_x$ to serve as the natural
conserved momentum.

For the ellipse, however, every bracket contains the
factor
\[
S(w)=a^2\sin^2 w + b^2\cos^2 w,
\]
which varies non-trivially along the orbit. Unlike $r_0$ in the
circular case, $S(w)$ is not a constant of motion and cannot be
absorbed into a rescaling of the canonical variables. As a result, there is no simple analogue of the explicit algebraic expression
for the conserved momentum $p_K$ (such as $L_z = x p_y - y p_x$ in the circle case),
whose square directly yields the Hamiltonian.

This $w$-dependence is the fundamental obstruction to
obtaining the elliptical operator algebra by extrapolation from the
circle. The remaining construction therefore requires a direct
operator ansatz, rather than a geometric guess borrowed from the
circular case.

\subsubsection{Operator Realization and Final Expressions}

Having obtained the full set of Dirac brackets for the elliptical constraint and imposed the parametrization
\[
x = a\cos w,\qquad y = b\sin w,
\]
the next step is to construct self-adjoint operators $\hat p_x$ and $\hat p_y$ acting on wavefunctions $\psi(w)$ that reproduce these brackets in the quantum theory.

Just like \cite{Scardicchio2002}, we use the general fact that for any sufficiently differentiable function \(F(w)\),$[d_w^n,\, F(w)]$ is a differential operator of order \(n-1\).  
Thus, commutators involving first–order differential operators naturally produce zeroth–order terms, matching the structure required by the Dirac brackets.  
This motivates the ansatz (ignoring $\hbar$)
\[
p_x = -i f(w)\frac{\partial}{\partial w} + A(w),
\qquad
p_y = -i g(w)\frac{\partial}{\partial w} + B(w),
\]
where \(f(w)\) and \(g(w)\) encode the kinematic weighting along the constrained curve, and \(A(w)\), \(B(w)\) represent possible ordering contributions.
\vspace{-0.2cm}

\subsubsubsection{Matching the Elementary Brackets}

The brackets (eq. \eqref{DiracBrac})
\[
[x,p_x] 
= \frac{i\,a^{2}\sin^{2}w}{a^{2}\sin^{2}w + b^{2}\cos^{2}w},
\qquad
[y,p_y] 
= \frac{i\,b^{2}\cos^{2}w}{a^{2}\sin^{2}w + b^{2}\cos^{2}w},
\]
fix the derivative coefficients uniquely:
\[
f(w) = -\frac{a\sin w}{a^{2}\sin^{2}w + b^{2}\cos^{2}w},
\qquad
g(w) = \frac{b\cos w}{a^{2}\sin^{2}w + b^{2}\cos^{2}w}.
\]
The mixed brackets $[x,p_y]$ and $[y,p_x]$ provide no new information for determining $A(w)$ or $B(w)$.

\subsubsubsection{Using the Commutator $[p_x,p_y]$}

The Dirac bracket yields
\[
[p_x,p_y] 
= 
-\frac{ab}{(a^{2}\sin^{2}w + b^{2}\cos^{2}w)^{2}}\frac{\partial}{\partial w}
+\frac{i(A(w)b\sin w - B(w)a\cos w)}
      {a^{2}\sin^{2}w + b^{2}\cos^{2}w}.
\]
Computing the commutator using the ansatz for $p_x$ and $p_y$ gives the same derivative term and produces a differential equation relating $A(w)$ and $B(w)$.

A second independent relation comes from enforcing the Heisenberg equations
\[
[x,H] = ip_x,\qquad [y,H] = ip_y,
\]
with $H = \tfrac12(p_x^2 + p_y^2)$.  
Using $[A,BC] = [A,B]C + B[A,C]$ and simplifying, one obtains
\begin{equation}
    \label{A in B}
A(w)
= \frac{B(w)\big(i2Sb^{2}\cos^{2}w - i2S^{2}\big) - a^{2}b\sin w}
       {2 i a b S \sin w \cos w},
\qquad
S = a^{2}\sin^{2}w + b^{2}\cos^{2}w.
\end{equation}

Substituting this into the equation from $[p_x,p_y]$ and solving (most efficiently by expressing $\sin w$ and $\cos w$ in exponential form) yields
\begin{equation}
    \label{Finally B}
B(w) 
= \frac{2iab\,e^{2iw}}
       {a^{2} - b^{2}
        + 2(-a^{2}-b^{2})e^{2iw}
        + (a^{2}-b^{2})e^{4iw}}
\;+\; C_{1},
\end{equation}
where $C_1$ is an integration constant determined by self-adjointness.  
The expression for $A(w)$ then follows from the previous relation.

\subsubsubsection{Final Form of the Momentum Operators}

The resulting operators are
\begin{equation}
    \label{px final}
p_x 
= -i\,\frac{a\sin w}{a^{2}\sin^{2}w + b^{2}\cos^{2}w}\,\frac{\partial}{\partial w}
\;+\; A(w),
\end{equation}
\begin{equation}
    \label{py final}
p_y 
= -i\,\frac{b\cos w}{a^{2}\sin^{2}w + b^{2}\cos^{2}w}\,\frac{\partial}{\partial w}
\;+\; B(w),
\end{equation}
with $A(w)$ and $B(w)$ as determined above, in \eqref{A in B} and \eqref{Finally B}.  
Together with
\[
x = a\cos w, \qquad y = b\sin w,
\]
these operators satisfy all Dirac-bracket commutation relations and therefore constitute the canonical quantization of a particle constrained to an ellipse.

\subsubsubsection{A Technical Remark on the Circle Derivation}

In the original analysis of the circle constraint \cite{Scardicchio2002}, an intermediate
relation of the form
\begin{equation}
    \label{oldeq}
a' \cos\theta + b' \sin\theta = -\,b \cos\theta + a \sin\theta
\end{equation}
was simplified by equating the coefficients of $\sin\theta$ and $\cos\theta$
separately. This step is justified only when the quantities multiplying these
functions are \emph{a priori} constant. In the constrained Hamiltonian problem,
however, $a(w)$ and $b(w)$ are dynamical expressions arising from the elimination
of redundant variables, and they cannot be assumed independent of~$\theta$.

Allowing $a(w)$ and $b(w)$ to vary admits additional solutions obtained by
redistributing terms between the sine and cosine channels (for example, rearranging as
$(a' + b)\cos\theta + (b' - a)\sin\theta = 0$ allows us to see that $a' + b = - sin \theta, b' - a = cos\theta$ still work, as opposed to $a' + b = 0, b'- a = 0$, which is what one gets by equating coefficients).
The inference in the original source therefore discards a family of admissible
solutions.

The derivation presented here does not rely on this assumption at any stage.
All expressions for $f(w)$, $g(w)$, $A(w)$, and $B(w)$ are obtained directly from the operator ansatz by using Dirac brackets and equations of motion, ensuring that the final operator
realizations are free of this algebraically crucial subtlety.
\vspace{-0.2cm}

\section{Conclusion}\label{sec4}

We have contrasted the quantization of a particle constrained to a circle with
that of a particle constrained to an ellipse, using both the standard
reparameterization approach and Dirac’s method. For the circle, the existence of
a natural angular coordinate and the familiar expression $L_z = xp_y - yp_x$
makes the system effectively one–dimensional and leads immediately to the
well-known operator realization of $\hat{p}_\theta$.

For the ellipse, the absence of a uniform parametrization forces the dynamics to
be expressed in terms of the arc–length coordinate $K$, involving an elliptic
integral. This non-uniformity appears directly in the Dirac brackets, whose
coefficients depend on the local geometry of the ellipse, and prevents a direct
analogue of the circular case. Several heuristics - such as attempting to
construct $p_K$ from the Hamiltonian, or modifying canonical brackets - provide
useful intuition but do not simplify the operator construction.

Using the Dirac brackets and a first-order differential operator ansatz, we
obtained explicit expressions for the momentum operators in equations \eqref{px final} and \eqref{py final},

These operators (as well as the intermediate steps, when the discussion around \eqref{oldeq} is appropriately handled) reproduce all Dirac brackets and reduce to the circle case when $a=b$.

Thus, while the circular system benefits from constant curvature and a well-studied
angular variable, the elliptical constraint leads to more complicated
quantum operators. The result provides a complete quantization of the system and
a concrete illustration of how even modest geometric generalizations can alter
the operator structure of constrained dynamics.

%\bibliographystyle{unsrt}
%\bibliography{references}

\end{document}